\documentclass[aps]{revtex4}
\usepackage{amssymb}
\usepackage{colordvi}
\usepackage{graphicx}
\usepackage{psfrag}

\newcommand{\lub}[1]{\mbox{$\;$* #1}}
\def\ssk{\smallskip}

\def\sec{\BrickRed} 
\def\emc{\Red} 
\def\boxc{\textOliveGreen} 
\def\textc{\textBlack} 

\newlength{\gnat} 
\begin{document}

{\footnotesize

\vspace*{-2\baselineskip}

\begin{minipage}[t]{.1\textwidth}
  \par\vspace{0pt}
  \includegraphics[width=1.8\textwidth]{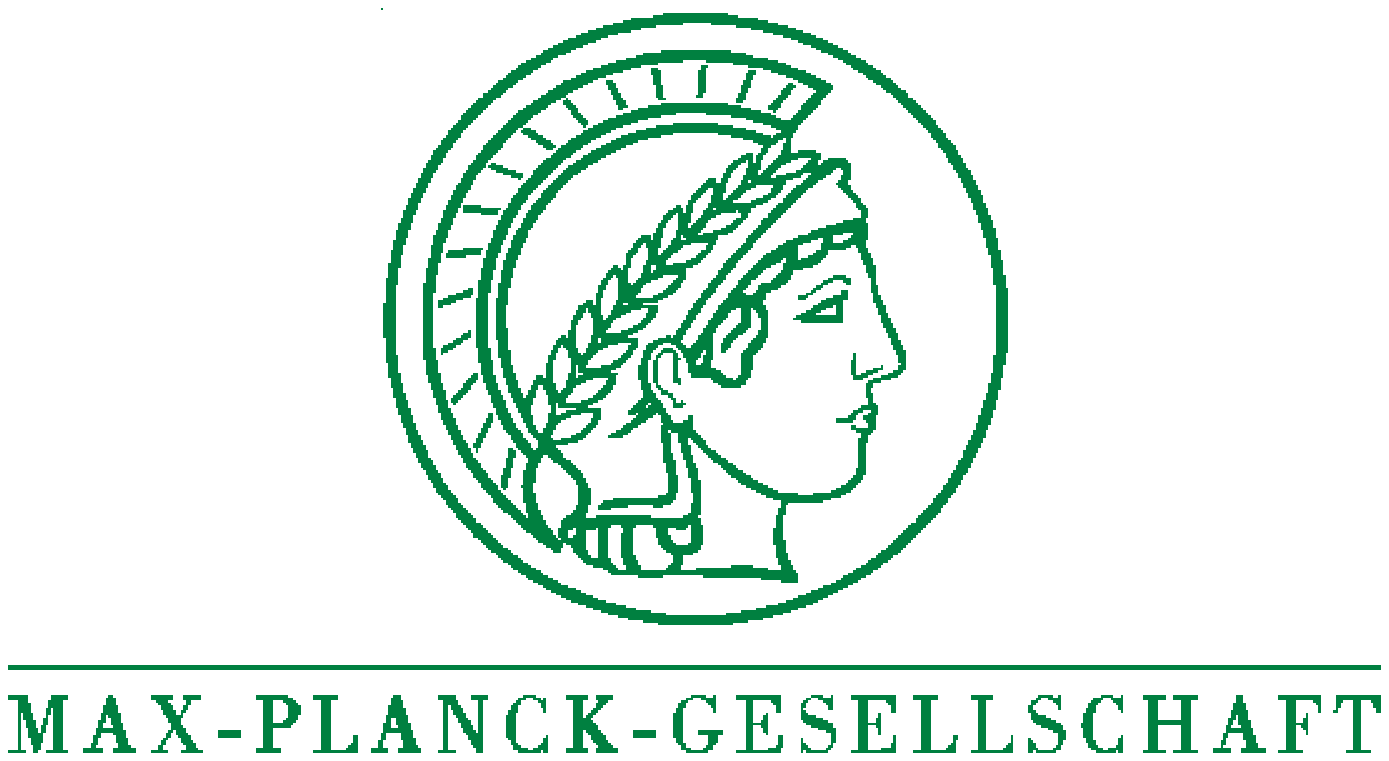}
\end{minipage}%
\begin{minipage}[t]{.8\textwidth}
  {\LARGE \Red{ 
      \centerline{Wall adsorption of a colloidal particle moving}
      \centerline{in a quiescent partially wetting fluid}}}
  
  \bigskip
 
  {\Large \centerline{\Blue{\it \underline{Alvaro Dom\'\i nguez}, Siegfried Dietrich}}}
  
  \bigskip
  
  {\large \BrickRed{\centerline{Max-Planck-Institut f\"ur Metallforschung \& ITAP--University of Stuttgart, Germany}}}
\end{minipage}%
\begin{minipage}[t]{.1\textwidth}
  \par\vspace{0pt}
  \includegraphics[width=\textwidth]{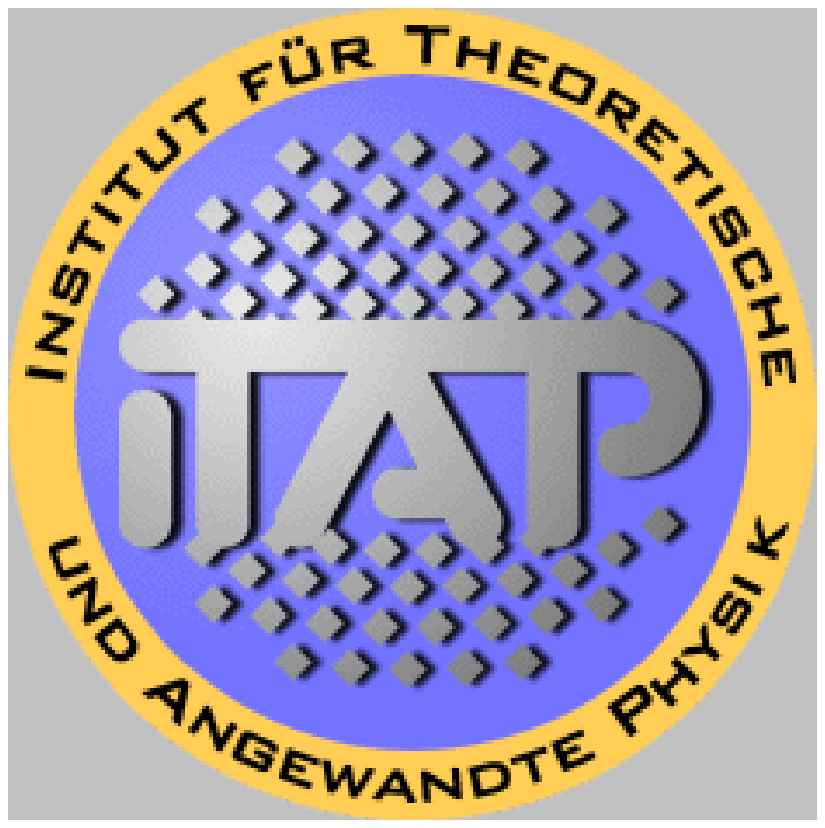}
\end{minipage}%

\bigskip

\boxc
\fbox{
  \textc
  \begin{minipage}[h]{\textwidth}
    In thermal equilibrium, a colloidal particle between two parallel
    plates immersed in a fluid which partially wets both the particle
    and the plates, is attracted by the walls. However, if the
    particle moves parallel to the plates, a hydrodynamic lift force
    away from the plates arises in the limit of low Reynolds number.
    We study theoretically the competition of these two effects and
    identify the range of velocity in which the velocity may serve as
    a parameter controlling the adsorption in microflows.
  \end{minipage}
  \boxc
  }
\textc


\boxc
\fbox{
  \textc
  \begin{minipage}[h]{\textwidth}

    {\large \centerline{\sec{Geometrical and physical setting}}}
    \smallskip
    \begin{minipage}[h]{\gnat}

      \begin{minipage}[t]{.45\textwidth}
        \par\vspace{0pt}
        \textBlue
        \psfrag{L=5R}{\Huge $\!\! L \approx (3$-$10) \, R$}
        \psfrag{h}{\Huge $\!\! h$}
        \psfrag{V}{\Huge $V$}
        \psfrag{R}{\Huge $\!\! R$}
        \psfrag{rho=rho}{\Huge $\varrho_{\rm fluid} \approx \varrho_{\rm particle}$}
        \resizebox{\textwidth}{!}{\includegraphics{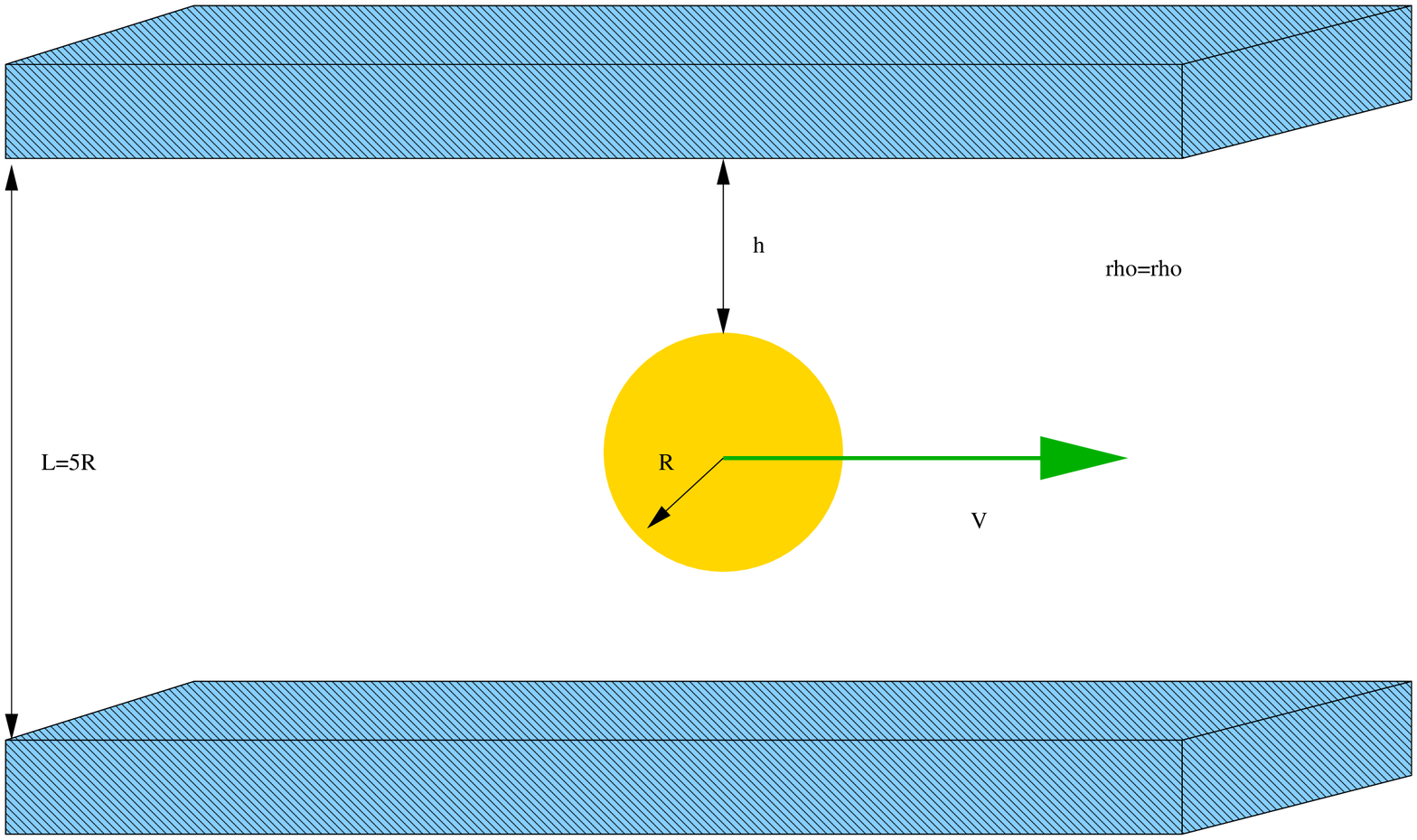}}
        \textc
      \end{minipage}\hfill
      \begin{minipage}[t]{.5\textwidth}
        \vspace{.03\textheight}
        A spherical particle moves \\
        in a quiescent fluid \\
        with constant velocity \\
        parallel to two infinite plates
      \end{minipage}

      \bigskip
      $\bullet$ Static interaction via dispersion force. Atomic potentials:
      \ssk

      \lub{Liquid-liquid, solid-solid: ${\cal V}_0 (r) = - 4 \varepsilon  (\sigma/r)^6 \qquad (\varepsilon \approx kT)$}
      \ssk

      \lub{Liquid-solid: ${\cal V}_{ls} (r) = \emc{A} {\cal V}_0(r)$}
      \ssk

      \mbox{\emc{Relative Hamaker constant}: $0 \leq A \leq 1$} \\
      \mbox{$A$ controls the wetting behavior. Contact angle: $\cos \theta \approx 2A -1$}

      \bigskip
      $\bullet$ Stationary incompressible flow:
      \medskip

      \centerline{\hfill$\nabla \cdot {\bf v} = 0$\hfill$\nabla^2 {\bf v} - \frac{R}{\mu V} \nabla (p+W_{\rm liquid}) = \emc{Re} \, ({\bf v} \cdot \nabla) {\bf v}$\hspace*{\fill}}
      
      \ssk

      \centerline{and no-slip boundary conditions}
      
      \medskip
      
      \mbox{Potential energy of the fluid in the dispersion field: $W_{\rm liquid}$}
      \ssk

      \mbox{\emc{Particle Reynolds number:} $Re = \varrho R V/\mu \lesssim 1$}

    \end{minipage}\hfill
    \begin{minipage}[h]{\gnat}
      
      \bigskip
      
      \mbox{$\bullet$ Effective potential energy of the particle, $W(h)=$}
      \medskip

      \hspace{-.025\textwidth}\begin{minipage}[t]{.5\textwidth}
        \vspace{.01\textheight}
        \hspace{.05\textwidth}\mbox{hydrodynamic+dispersion $=$}
        \vspace{.005\textheight}
        \begin{displaymath}
          kT \, \frac{R}{\sigma} \, Re^2 \; w_1 \left( \frac{h}{R}, \frac{L}{R}, Re \right) - \mbox{}
         \end{displaymath}
         \begin{displaymath}
           \mbox{} - (1-A) \, \varepsilon \; w_2 \left( \frac{h}{R}, \frac{L}{R} \right)
      \end{displaymath}
      \end{minipage}\hspace{.02\textwidth}
      \begin{minipage}[t]{.5\textwidth}
        \par\vspace{0pt}
        \psfrag{W}{$W$}
        \psfrag{h}{$h/R$}
        \psfrag{largeRe}{\textBlue $(R/\sigma) Re^2 \gg 1-A$}
        \psfrag{smallRe}{\textGreen $(R/\sigma) Re^2 \ll 1-A$}
        \includegraphics[width=\textwidth]{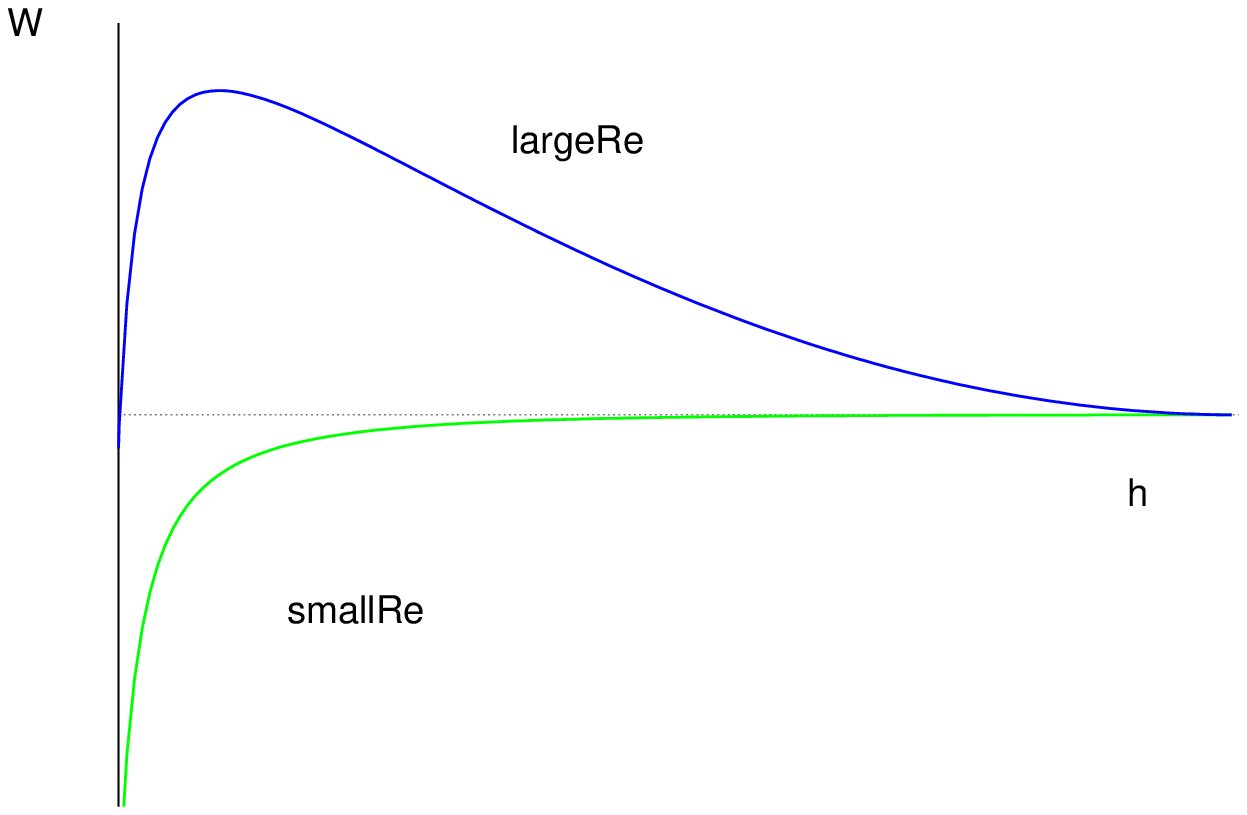}
        \textc
      \end{minipage}

      
      $\bullet$ \mbox{Motion in the $h$-direction: Brownian + relaxational dynamics:}
      \begin{displaymath}
        6 \pi \mu R v \lambda(h) + \frac{d W}{d h} = 0 , \qquad 
        \lambda(h) = \textrm{wall-correction to Stokes drag}
      \end{displaymath}
      
      $\bullet$ \mbox{The Fokker-Planck equation for the particle motion yields}\\
      \mbox{$\;\;\;$the mean first-passage time, ${\cal T}$, from the midplane, $h=L/2-R$,}\\
      \mbox{$\;\;\;$to wall contact, $h \approx \sigma$:}
      \medskip

      \begin{displaymath}
        {\cal T} \propto \int_{L/2-R}^\sigma \!\!\!\!\!\!\!\!\! {\rm d} y \;\; \int_{L/2-R}^y \!\!\!\!\!\!\!\!\! {\rm d} z 
        \;\; \lambda(y) \; \exp{[W(y)-W(z)]/kT}
      \end{displaymath}

      $\bullet$ \mbox{${\cal T}$ is normalized to ${\cal T}_{\rm free} =$
        passage time in the absence of forcing:\textRed} \\
      \hspace*{.1\textwidth}\begin{minipage}[t]{.6\textwidth}
          ${\cal T} < {\cal T}_{\rm free} \Rightarrow$ enhanced adsorption \\
          ${\cal T} > {\cal T}_{\rm free} \Rightarrow$ delayed adsorption
        \end{minipage}
      \end{minipage}\hspace*{\fill}

      \textc
  \end{minipage}
  \boxc
  }
\textc


\boxc
\fbox{
  \textc
  \begin{minipage}[h]{\textwidth}
    {\large \centerline{\sec{Results}}}
    \bigskip

    \begin{minipage}[h]{\gnat}
      $\bullet$ 
      \begin{minipage}[t]{.95\textwidth}
        We study ${\cal T}$ as a function of $A$, $Re$ and $R$
        for values of the \\
        parameters $\sigma$, $\varrho$, $\varepsilon$, $\mu$
        appropriate for atomic liquids
      \end{minipage}
      \medskip

      \begin{minipage}[t]{.5\textwidth}
        \par\vspace{0pt}
        \psfrag{T}{\Large $\!\!{\cal T}/{\cal T}_{\rm free}$}
        \psfrag{Re}{\Large $Re$}
        \psfrag{A=0}{\Large $\!\!\!\!\!\!\!\!\!\!\!A= \Red{0}, \Green{0.5}, \Blue{0.8}, \Magenta{0.95}$}
        \psfrag{R=0}{\Large $R/\sigma = 10^3$}
        \psfrag{1e+04}{\Large $\;\;\; 10^4$}
        \psfrag{1e+03}{$\,$}
        \psfrag{1e+02}{\Large $\;\;\; 10^2$}
        \psfrag{1e+01}{$\,$}
        \psfrag{1e+00}{\Large $\;\;\;\; 1$}
        \psfrag{1e-01}{$\,$}
        \psfrag{1e-02}{\Large $\;\; 10^{-2}$}
        \psfrag{0}{\Large $0$}
        \psfrag{0.04}{\Large $0.04$}
        \psfrag{0.08}{\Large $0.08$}
        \psfrag{0.12}{\Large $0.12$}
        \psfrag{0.16}{\Large $0.16$}
        \psfrag{0.2}{\Large $0.2$}
        \resizebox{\textwidth}{!}{\includegraphics{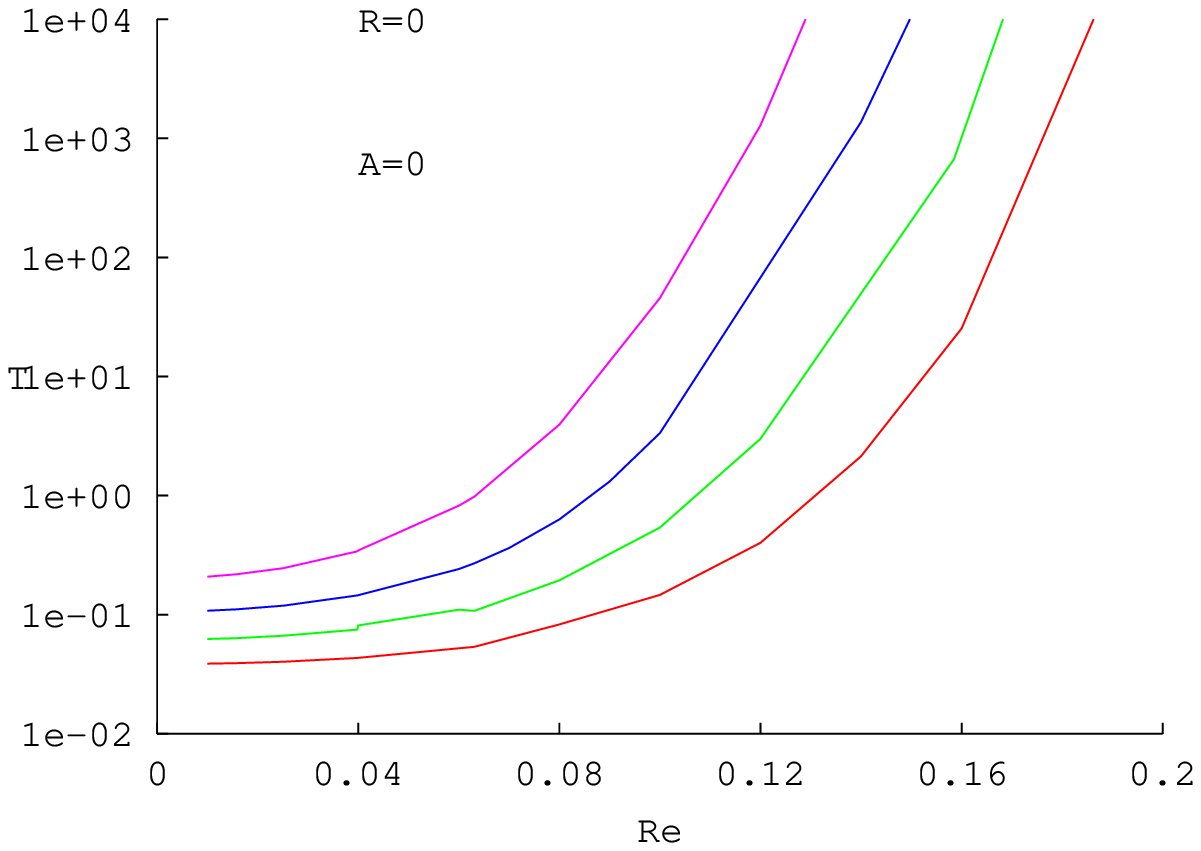}}
        \textc
      \end{minipage}%
      \begin{minipage}[t]{.5\textwidth}
        \par\vspace{0pt}
        \psfrag{T}{\Large $\!\!{\cal T}/{\cal T}_{\rm free}$}
        \psfrag{Re}{\Large $Re$}
        \psfrag{A=0}{\Large $\!\!\!\!\!\!\!\!\!\!\!A= \Red{0}, \Green{0.5}, \Blue{0.8}, \Magenta{0.95}$}
        \psfrag{R=0}{\Large $R/\sigma = 10$}
        \psfrag{1e+02}{\Large $\;\;\; 10^2$}
        \psfrag{1e+01}{\Large $\;\;\; 10^1$}
        \psfrag{1e+00}{\Large $\;\;\;\; 1$}
        \psfrag{1e-01}{\Large $\;\; 10^{-1}$}
        \psfrag{0}{\Large $0$}
        \psfrag{0.2}{\Large $0.2$}
        \psfrag{0.4}{\Large $0.4$}
        \psfrag{0.6}{\Large $0.6$}
        \psfrag{0.8}{\Large $0.8$}
        \psfrag{1}{\Large $1.0$}
        \psfrag{1.2}{\Large $1.2$}
        \resizebox{\textwidth}{!}{\includegraphics{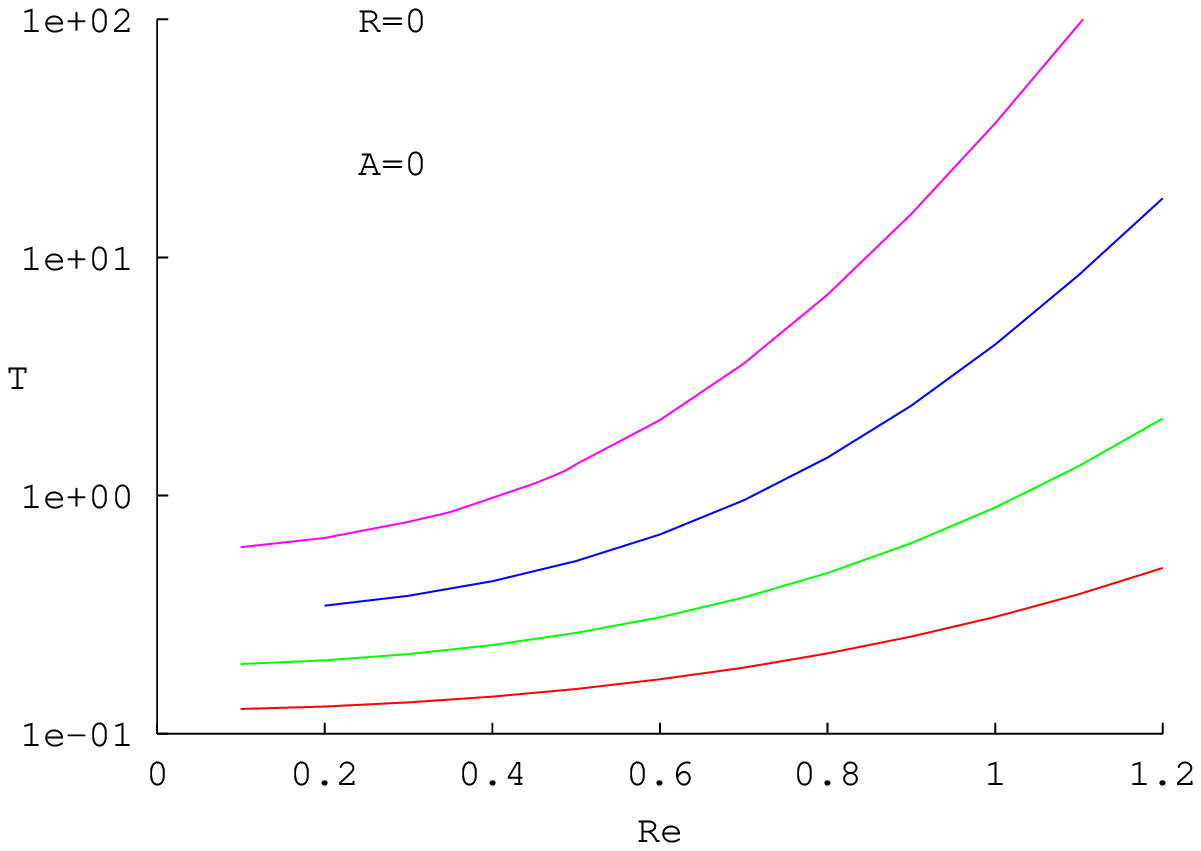}}
      \end{minipage}%
      
      $\bullet$ \mbox{The qualitative shape of the function ${\cal T} (A, Re \lesssim 1)$ is} \\
      \mbox{$\;\;\;$ independent of $R$, $L$ in the range}
      \ssk

      \centerline{$10 \lesssim R/\sigma \lesssim 10^4 \qquad 3 \lesssim L/R \lesssim 10$}
      
    \end{minipage}\hfill
    \begin{minipage}[h]{\gnat}
      $\bullet$ \mbox{The sensitivity of ${\cal T}$ to changes in $Re$ increases with $Re$ and $R/\sigma$}
      \medskip

      $\bullet$ \mbox{Define a critical $Re_c$ by the condition ${\cal T} = {\cal T}_{\rm free}$:}

      \begin{minipage}[t]{.5\textwidth}
        \par\vspace{0pt}
        \psfrag{log Re}{\Large $\!\!\!\log Re_c$}
        \psfrag{log (1-A)}{\Large $\log (1-A)$}        
        \psfrag{slope=1/4}{\Large slope$=1/4$}
        \psfrag{R=0}{\Large $R/\sigma = \Magenta{10}, \Blue{10^2}, \Green{10^3}, \Red{10^4}$}
        \psfrag{0}{\Large $0$}
        \psfrag{-0.2}{\Large $-0.2$}
        \psfrag{-0.4}{\Large $-0.4$}
        \psfrag{-0.6}{\Large $-0.6$}
        \psfrag{-0.8}{\Large $-0.8$}
        \psfrag{-1}{\Large $-1.0$}
        \psfrag{-1.2}{\Large $-1.2$}
        \psfrag{-1.4}{\Large $-1.4$}
        \psfrag{-1.6}{\Large $-1.6$}
        \resizebox{\textwidth}{!}{\includegraphics{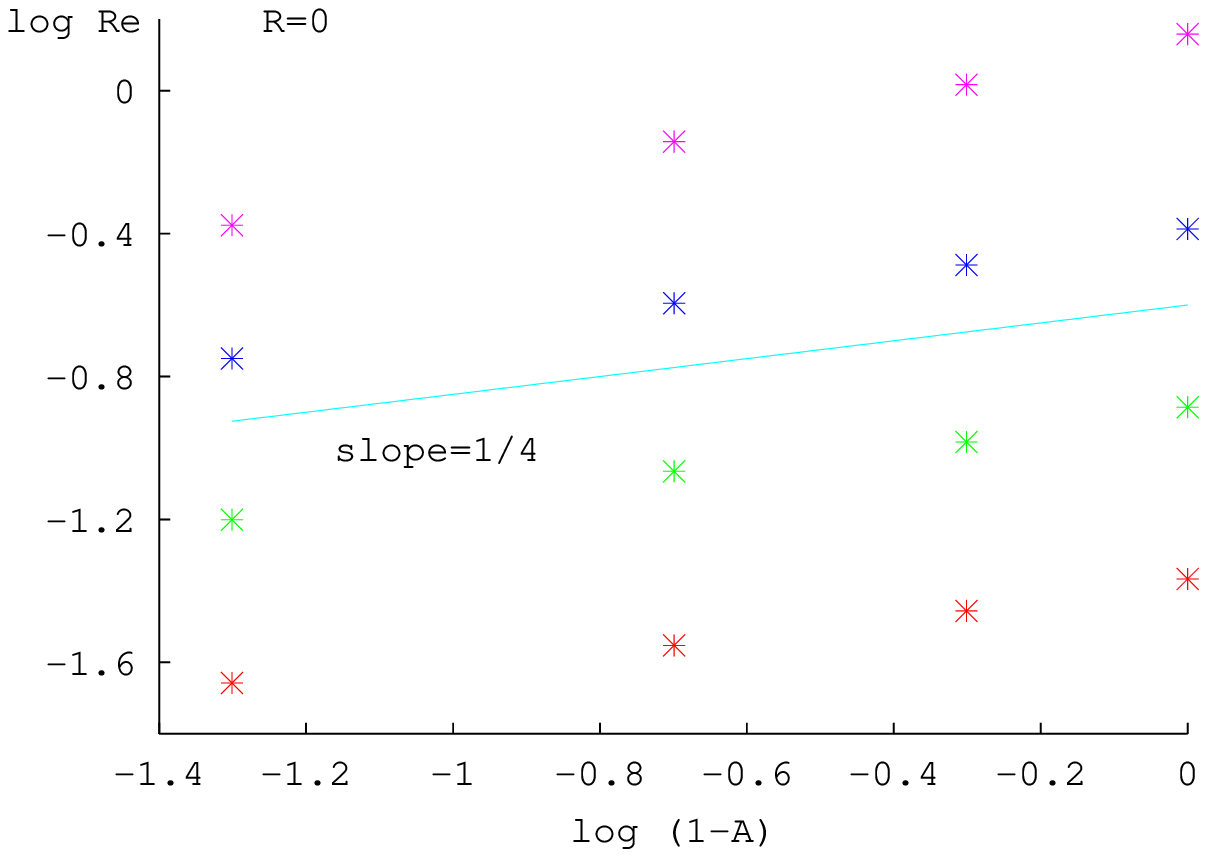}}
      \end{minipage}%
      \begin{minipage}[t]{.5\textwidth}
        \bigskip
        \begin{center}
          Collapse on a master curve:
          \bigskip

          $(R/\sigma) Re_c^2 \propto \sqrt{1-A}$
          \bigskip
          
          with departures at $R/\sigma \sim 10$
        \end{center}
      \end{minipage}

      $\bullet$ \mbox{Some numbers: for atomic liquids $V ($m$\,$s$^{-1}) \sim Re / R (\mu$m)}
      \ssk
      
      \centerline{$\;\;\;$ $R \sim 1 \, \mu$m $\Rightarrow Re_c \sim 0.04 \Rightarrow V_c \sim 4 \,$cm$\,$s$^{-1}$}
    \end{minipage}\hspace*{\fill}
    
  \end{minipage}
  \boxc
  }
\textc

\boxc
\fbox{
  \textc
  \begin{minipage}[h]{\textwidth}
    {\large \centerline{\sec{Conclusions}}}
    \medskip
    
    $\bullet$ In the \emc{$\mu$-meter scale}, the time of adsorption by the
    wall can be controlled in a realistic range of values of the
    particle velocity.
    
    $\bullet$ In the \emc{nano-meter regime}, the theory presented is
    pushed to its limits and suggests that the hydrodynamic lift is
    irrelevant.
    
    $\bullet$ In particular, hydrodynamic lift is unlikely to explain
    the results of the MD simulation by Drazer~et~al., PRL {\bf 89}
    (2002) 244501.

    \end{minipage}
  \boxc
  }
\textc


}

\end{document}